\documentstyle[preprint,aps]{revtex}
\tighten
\begin{document}
\draft
\preprint{\hfill NCL93--TP8}
\title{Finite temperature effective actions for double wells
}
\author{Ian Moss}
\address{
Department of Physics, University of Newcastle Upon Tyne, NE1 7RU U.K.
}
\date{September 30, 1993}
\maketitle
\begin{abstract}
The problem of calculating the effective potential for a real scalar
field with a double--well potential is a possible preliminary to
understanding symmetry breaking phase transitions in the early
universe. It is argued here that it is necessary to include multiple
saddle points in the path integral formalism and that quadratic source
terms are important. The resulting effective action has a well--defined
propagator expansion with a positive mass and gives a new way of
understanding high temperature expansions, but the results are equally
valid at low or zero temperatures.
\end{abstract}
\pacs{03.70.+k,98.80.Cq}
\narrowtext

\section{INTRODUCTION}

The favoured models of particle physics and cosmology imply that the
hot material of the early universe underwent a series of phase
transtions where the nature of the fundamental forces changed. These
transitions are associated with the breaking of a symmetry of the
dynamical equations due to a Higgs field developing an expectation
value in the lower temperature phase.

The physics of these high temperature phase transitions is often
described by an effective potential for the expectation value of the
Higgs field \cite{goldstone,dolan,weinberg,linde1}. The definition of this
potential for a real scalar field $\phi$ and the symmetry group $Z_2$
is the issue with which we shall be concerned in what follows.

In order to break the symmetry the classical potential should have a
global minimum away from the origin, as shown in figure \ref{fig1}.
In applications concerning the early universe we are particularly
concerned with the way in which the field leaves the origin and
approaches the global minimum. This can be described by using a quantum
effective potential $V_\beta(\phi)$, but some care has to be taken
in defining it. A simple loop expansion fails because the
mass is imaginary at the classical level, and the usual proceedure
would be to use some form of `corrected' mass, related to resummation
of subsets of the graphs in the loop expansion.

The definition of the quantum effective potential $V_\beta(\phi)$ is
even more difficult in a system which has a finite volume, a situation
which arises when cosmological effects are important. Then
$V_\beta(\phi)$ can be shown to be convex \cite{symanzik}. The underlying
reason for the convexity is that the ground state is a superposition
of wave functions $\Psi_\pm$ that are peaked about the homogeneous
field values $\phi_\pm$ at the minima of the potential. In a path integral
construction of the effective potential there are two equally dominant
field configurations \cite{rivers}. When $V_\beta(0)$ is evaluated it
is dominated by both minima of the classical potential and tells us
nothing about the behaviour of the system near to the symmetric point
$\phi=0$.

Some years ago it was suggested that a more satisfactory approach to
the finite temperature effective action \cite{hawking1,moss} was to
follow Cornwall, Jackiw and Tomboulis \cite{cornwall} and introduce a
source $K(x,x')$ coupled to quadratic terms in the action. The reason
for the effectiveness of this source is that it is able to prepare the
quantum state of the field to be concentrated near to the symmetric
value, the place it would occupy before the onset of a supercooled
phase transition. This would be similar to using a corrected mass in the
usual approach, but would apply equally to extreme supercooling
where $T=0$ as to high temperatures.

In the formalism of reference \cite{cornwall}, the
effective action is a function of the field expectation value $\phi(x)$
and the connected propagator $G(x,x')$. The physical solutions satisfy
\begin{eqnarray}
{\delta\Gamma[\phi,G]\over\delta\phi(x)}&&=
J(x)-\int d\mu(x')K(x,x')\phi(x'),\label{j}\\
{\delta\Gamma[\phi,G]\over\delta G(x,x')}&&=
-\case1/2K(x,x').\label{k}
\end{eqnarray}

If the quadratic source is set to zero then we can solve the second
equation for $G[\phi]$. The result would be the
usual effective action,
\begin{equation}
\Gamma[\phi]=\Gamma[\phi,G[\phi]].
\end{equation}
Amelino--Camelia and So--Young Pi \cite{amelino} have shown that this
approach provides an efficient way of resuming the loop diagrams to
obtain the corrected mass.

The new idea that will be used here is that
other linear combinations of \ref{j} and \ref{k} can be solved for
$G[\phi]$ to obtain different effective actions
$\Gamma[\phi,G[\phi]]$. From the different possibilities, one is
chosen that satisfies the following criteria:

1. The effective action generates effective field equations,
\begin{equation}
{\delta\Gamma[\phi]\over\delta\phi}=0
\end{equation}

2. It has a well--defined loop expansion.

3. It reduces to the calssical action as $\hbar\to 0$,
\begin{equation}
\Gamma[\phi]=I[\phi]+O(\hbar)
\end{equation}

When the  mass is imaginary it is possible to obtain an effective
potential this way which has a symmetry breaking minimum in the
infinite volume limit. The field can then evolve smoothly or tunnel
through the potential from the symmetric point. At high temperatures
the potential is similar to the usual potential with a corrected mass,
but the method gives a uniform treatment of low as well as high
temperatures.

In constructing the effective action for finite volumes it is necessary
to allow several
field configurations to dominate the path integral simultaneously. Near
to the symmetric value of the field at $T<T_c$ any linear source terms
become very small and the quadratic source terms are the most
important. There is some similarity here to work by Laurie
\cite{laurie} and Cahill \cite{cahill}, who use quadratic sources but
do not take into account the effects of multiple saddle points. Beyond
the minima of the potential, the linear sources take over and the
potential reverts to the usual form.

The inverse of $G(x,x')$ is the quantum fluctuation operator for the
Higgs fields. With the construction described here the operator comes
out to be positive definite. Furthermore, the potential has a barrier
around the origin which has a different height than for some versions
of the temperature corrected potential (e.g. \cite{dolan}), but since
differences can arise from resumming different sets of graphs these
differences may not be significant. For high temperature field theory
the new approach can be viewed as an attempt to justify the use of
corrected masses by introducing a quadratic source in the path integral.

\section{DOUBLE WELLS AT ZERO TEMPERATURE}

This section begins by setting up the definition of the effective
action as a function of composite fields following reference
\cite{cornwall}. We then go on to consider the modifications that are
necessary when there is more than one field configuration dominating
the path integral. A Riemannian signature (++++) will be used and units
in which the velocity of light $c=1$.

The classical action for a real scalar field will be taken to be of the
form
\begin{equation}
I[\phi]=\int
d\mu(x)\left\{-\case1/2\nabla\phi.\nabla\phi+V(\phi)\right\},
\end{equation}
with a classical double well potential
\begin{equation}
V(\phi)=-\case1/2\mu^2\phi^2+\case1/4\lambda\phi^2.
\end{equation}

The generating function $Z[J,K]$ is defined by a path integral,
\begin{equation}
Z[J,K]=\int d\mu[\phi]e^{-I_{JK}[\phi]/\hbar}
\end{equation}
where
\begin{eqnarray}
I_{JK}[\phi]=&&I[\phi]+\int d\mu(x)J(x)\phi(x)+\nonumber\\
&&\case1/2\int d\mu(x,x')\phi(x)K(x,x')\phi(x')
\end{eqnarray}

If we set $W=-\hbar\ln Z$, then the expactation values $\phi$ and
connected propagators $G$ are
\begin{eqnarray}
\phi(x)&=&{\delta W\over\delta J(x)}\\
G(x,x')&=&2{\delta W\over\delta K(x,x')}-
{\delta W\over\delta J(x)}{\delta W\over\delta J(x')}\label{g}
\end{eqnarray}

The effective action is defined by eliminating $J$ and $K$ in favour of
$\phi$ and $G$,
\begin{eqnarray}
&&\Gamma[\phi,G]=W[J,K]-\int d\mu(x)\phi(x)J(x)\nonumber\\
&&-\case1/2\int d\mu(x,x')\phi(x)K(x,x')\phi(x')
-\case1/2\int d\mu(x,x')G(x,x')K(x,x').\label{gamma}
\end{eqnarray}

Before proceeding further it should be instructive to examine the
nature of the field configurations $\phi_s$ that dominate the path
integral. This can be simplified to the case of homogeneous fields, for
which
\begin{equation}
J(x)=j\hbox{,~~~}K(x,x')=k\delta(x-x')
\end{equation}
with constants $j$ and $k$. The saddle point condition,
\begin{equation}
{\delta I_{JK}\over\delta\phi}=0\label{saddle},
\end{equation}
then implies that
\begin{equation}
V'(\phi_s)+j+k\phi_s=0.
\end{equation}
Solutions are sketched in figure \ref{fig2}. There are three saddle
points when $j^2<4(\mu^2-k)^3/27\lambda$, one of which has a much
larger action than the other two and can usually be neglected. There is
a series expansion for the solutions in powers of $j$,
\begin{equation}
\phi_{s\pm}=\pm{(\mu^2-k)^{1/2}\over\sqrt{\lambda}}-
\case1/2(\mu^2-k)^{-1}j+O(j^2)\label{exp}.
\end{equation}

The generating function $Z$ can be evaluated by shifting the fields and
constructing an asymptotic expansion in $\hbar$ around each of the
saddle points seperately. The two series have then to be summed,
\begin{equation}
Z\sim e^{-W_+/\hbar}+e^{-W_-/\hbar}\label{z}.
\end{equation}
If we set $\phi=\phi_\pm+\tilde\phi$, then the asymptotic expansions
are generated by
\begin{eqnarray}
&&e^{-W_\pm[J,K]/\hbar}=e^{-I_{JK}[\phi_\pm]/\hbar}\nonumber\\
&&\int d\mu[\tilde\phi]
\exp{1\over\hbar}\left\{-\int\tilde\phi(\Delta_\pm+K)\tilde\phi-\int
J_{int}\tilde\phi
-I_{int}\right\}.\nonumber\\
\end{eqnarray}
The interaction term $I_{int}$ consists of the terms from
$I[\phi_\pm+\tilde\phi]$ that are cubic or higher order in
$\tilde\phi$. The other terms are,
\begin{eqnarray}
J_{int}(x)&=&{\delta I_{JK}\over\delta\phi(x)}\label{jint}\\
\Delta(x,x')&=&{\delta^2 I\over\delta\phi(x)\delta\phi(x')}\label{dint}
\end{eqnarray}

Instead of choosing solutions of the classical equations for $\phi_\pm$
it is better to use quantum corrected background fields. Define local
quantum expectation values of the field operator by
\begin{equation}
\langle\phi\rangle_\pm={\delta W_\pm\over\delta J},
\end{equation}
then the fields $\phi_\pm$ will be defined by
\begin{equation}
\phi_\pm=\langle\phi\rangle_\pm\label{phic}.
\end{equation}
Because of \ref{saddle}, $\phi_\pm=\phi_{s\pm}$ to leading order in
$\hbar$.

The expectation value of the field $\phi$ is given by the generating
function,
\begin{equation}
\phi=Z^{-1}{\delta Z\over\delta J}.
\end{equation}
{}From the asymptotic expansion \ref{z},
\begin{equation}
\phi\sim {e^{-W_+/\hbar}\over Z}\phi_++{e^{-W_-/\hbar}\over Z}\phi_-
\end{equation}
Some re--arrangement allows this to be written as
\begin{equation}
\phi\sim\case1/2(\phi_++\phi_-)-
\case1/2(\phi_+-\phi_-)\,\Theta,\label{phi}
\end{equation}
where
\begin{equation}
\Theta=\tanh\left({W_+-W_-\over2\hbar}\right).
\end{equation}

The propagator can be expressed likewise. We set
\begin{equation}
\hbar G_\pm(x,x')=\langle\tilde\phi(x)\tilde\phi(x')\rangle_\pm
\label{gc}.
\end{equation}
{}From the definition of $G$, equation \ref{g}, we get
\begin{eqnarray}
&&\hbar G(x,x')\sim {e^{-W_+/\hbar}\over Z}\hbar G_+(x,x')
+{e^{-W_-/\hbar}\over Z}\hbar G_-(x,x')+\nonumber\\
&&{e^{-(W_++W_-)/\hbar}\over Z^2}(\phi_+(x)-\phi_-(x))
(\phi_+(x')-\phi_-(x')).\nonumber\\
\end{eqnarray}
We write this in the form,
\begin{equation}
\hbar G\sim\case\hbar/2(G_++G_-)-
\case\hbar/2(G_+-G_-)\,\Theta+\rho^2,\label{GG}
\end{equation}
where\FL
\begin{equation}
\rho^2(x,x')=\case1/4(1-\Theta^2)(\phi_+(x)-\phi_-(x))
(\phi_+(x')-\phi_-(x')) \label{rho}.
\end{equation}

The meaning of these expressions can be made clearer by returning to
the homogeneous case. To leading order in $\hbar$ we replace $\phi_\pm$
in equation \ref{phi} by the classical solution $\phi_{s\pm}$ (equation
\ref{exp}),
\begin{equation}
\phi=-\phi_0\tanh(\Omega\phi_0j/\hbar)-\case1/2(\mu^2-k)^{-1}j
+O(j^2)\label{p}
\end{equation}
where $\phi_0(k)$ is the value of the classical solution at $j=0$ and
$\Omega$ is the spacetime volume. Evidently, $\phi$ covers the whole
range of values from $-\phi_0$ to $\phi_0$ over a very small range of
$j$. This is the first indication that we should set $j\to0$ as
$\Omega\to\infty$.

In the same approximation, the propagator contains the term $\rho$
given by,
\begin{equation}
\rho^2=(1-\Theta^2)\phi_0^2+O(j^2)=\phi_0^2-\phi^2+O(j)\label{r}
\end{equation}
If we chose a form for $G$ where $\rho=0$ then we must have
$\phi\sim\phi_0$, i.e. the value of $\phi$ is set by the quadratic
source. (This condition on $\rho$ was implicitly assumed in ref
\cite{amelino}, although it holds automatically only as long as the
mass is real.)

We turn now to the effective action. For the double saddle \ref{z},
\begin{equation}
-\hbar\ln
Z\sim\case1/2(W_++W_-)+\case1/2\hbar\ln(1-\Theta^2)\label{lnz}
\end{equation}
We can construct an asymptotic series for $W_+$ in terms of Feynmann
diagrams with propagator $G_0=(\Delta+K)^{-1}$. The contributions that
are two--loop and higher will be labelled $W_+^{(2)}$,
\begin{equation}
W_+\sim I_{JK}[\phi_+]+\case1/2\hbar\ln\det(\Delta_++K)+W_+^{(2)}.
\end{equation}

Two conditions will be imposed, prompted by the discussion of the
homogeneous case. First of all, the condition that $\rho\to0$ as
$\Omega\to\infty$ will be imposed on $G$. This guarantees that the
propagator has a simple local form to leading order in $\hbar$. It also
forces the state of the system away from the global minima of the
potential when $\phi$ is small. It follows from equation \ref{rho} that
$\phi_+\to\phi_-$ or $\Theta\to\pm1$. The first of these would
correspond to the situation which applies to single well potentials. In
the second case, the condition forces only one of the two saddles to
dominate. If $\Theta\to-1$, then it follows from \ref{phi} that
$\phi\to\phi_+$, from \ref{GG} that $G\to G_+$ and from \ref{p} that
$\Omega J\to-\infty$.

The second condition is that $J\to0$ as $\Omega\to\infty$ (taking
$J\sim\Omega^{-1/2}$ for example). This condition allows us to
solve for $K[\phi]$, or equivalently $G[\phi]$. If we chose
$K\to0$ instead, then $G$ would be a propagator with an imaginary
mass in the limit $\hbar\to0$, leading to problems with the effective
action at the one--loop level. These problems could be solved by
resumming series of graphs, but this is a proceedure which the present
approach seeks to avoid. The significance of the two conditions is
explained further in the next section.

With these conditions and equation \ref{gamma} we get,
\begin{equation}
\Gamma[\phi]\sim I[\phi]-\case{\hbar}/2{\rm tr}(GK)
+\case\hbar/2\ln\det(\Delta+K)+W^{(2)}\label{res}.
\end{equation}
The source $K$ is found by the solution of equation \ref{phic} and $G$
is given by equation \ref{gc}. Diagramatically, equation \ref{phic} is
equivalent to setting the sum of all tadpole diagrams in figure
\ref{fig4} to zero. The diagrammatic expansion of the propagator is
shown in figure \ref{fig5}.

The effective action can be rewritten also in a more elegant form.
First, we separate the one--loop and higher part $G^{(1)}$ of $G$,
\begin{equation}
G=G_0+G^{(1)}.
\end{equation}
Then,
\begin{equation}
G_0=G(1-G^{-1}G^{(1)})
\end{equation}
and
\begin{equation}
K=G_0^{-1}-\Delta=(1-G^{-1}G^{(1)})^{-1}G^{-1}-\Delta.
\end{equation}
Whan substituted into equation \ref{res}, The propagator corrections
have the effect of cancelling all of the two--particle reducible
diagrams in $W^{(2)}$. This allows the result to be written in the
form,
\begin{equation}
\Gamma[\phi]\sim I[\phi]-\case{\hbar}/2{\rm tr}(1-G\Delta)
-\case\hbar/2\ln\det G+\Gamma^{(2)}\label{main},
\end{equation}
where $\Gamma^{(2)}$ has only two--particle ireducible diagrams in the
propagator $G$. The result now looks the same as the one in reference
\cite{cornwall}, but it should be realised that the result only takes
this simple form in the limit that $J\to0$.

To leading order in $\hbar$, $G=(\Delta+K)^{-1}$ and $K$ is given by
$J_{int}=0$ (\ref{jint}). Therefore, in the homogeneous case,
\begin{equation}
k=\mu^2-\lambda\phi^2
\end{equation}
and from equation \ref{dint},
\begin{equation}
G^{-1}=-\nabla^2+V''(\phi)+k=-\nabla^2+2\lambda\phi^2.
\end{equation}
There are two remarkable features of this result. The first is that the
operator is positive definite even when $V''(\phi)$ is negative. The
second is that it differs substantially from the corresponding operator
for a single well,
\begin{equation}
G^{-1}=-\nabla^2+V''(\phi)=-\nabla^2-\mu^2+3\lambda\phi^2.
\end{equation}
where $\mu$ is imaginary.

\section{A DOUBLE--POTENTIAL FOR FINITE VOLUMES}

The form of the effective action that has been arrived at depends
crucially upon the conditions that where imposed on $G$. It would be
instructive to investigate these conditions further by keeping $\rho$
non--zero. Homogeneous fields will be used. Equation \ref{r} for $\rho$
tell us that
\begin{equation}
\phi_\pm=\pm(\rho^2+\phi^2)^{1/2}+O(j).
\end{equation}
When substituted into equation \ref{lnz} for $\ln Z$ this gives
\begin{equation}
-\hbar\ln Z=I_{JK}[\sqrt{\rho^2+\phi^2}]
+\case1/2\hbar\ln(1-\Theta^2)+O(j^2),
\end{equation}
with $O(j^2)$ because the action is stationary at $\phi_+$. From the
definition of the effective action, we have
\begin{eqnarray}
\Gamma=&&I_{JK}[\sqrt{\rho^2+\phi^2}]
+\case1/2\hbar\ln(1-\Theta^2)\nonumber\\
&&-\Omega j\phi-\case1/2\Omega k(\phi^2+\rho^2)+O(j^2).
\end{eqnarray}
Now we use
\begin{equation}
j={\hbar\over\phi_0\Omega}\tanh^{-1}\Theta+O(j^2)
\end{equation}
from equation \ref{p} and
\begin{equation}
\Theta=-{\phi\over\sqrt{\rho^2+\phi^2}}+O(j)
\end{equation}
from equation \ref{r}. We define a `zero--loop' effective potential by
$V^{(0)}=\Gamma/\Omega$, and then
\begin{equation}
V^{(0)}(\rho,\phi)=V(\sqrt{\rho^2+\phi^2})+V_\Omega(\rho,\phi)+O(j)
\end{equation}
where
\begin{equation}
V_\Omega(\rho,\phi)=
{\hbar\over\Omega}\left\{{\phi\over\sqrt{\rho^2+\phi^2}}
\tanh^{-1}{\phi\over\sqrt{\rho^2+\phi^2}}
+\case1/2\ln{\rho^2\over\rho^2+\phi^2}\right\}.
\end{equation}
This double--potential has been plotted in figure \ref{fig6}.

The global minimum of the double--potential lies at $\phi=0$ and
$\rho=\phi_0$, where $\phi_0^2=\mu^2/\lambda$. This minimum represents
a superposition of states at the two minima $\pm\phi_0$. However, as
$\Omega\to\infty$ the mixing between the two minima approaches zero and
this is represented by the valley bottom of the potential levelling off
allowing the field to remain at any point in the bottom of the valley
indefinitely.

Now consider what happens when the field is released from the vicinity
of the symmetric point. If $\phi$ and $\rho$ are very small, then the
potential is steepest in the $\rho$ direction and therefore $\rho$
starts to grow. This is the  growth in $\langle\phi^2\rangle$ that has
been noticed before in the context of early universe phase transitions
\cite{hawking1}.

The subsequent evolution of $\phi$ depends on the initial `velocity'.
If the initial component of velocity in the $\rho$ direction is
negligeable and the volume is very large, then the growth in $\rho$
will be tiny and $\rho=0$ will be a good approximation. Equations for
the evolution of $\phi$ can then be found from variation of the
effective action. In the previous section we saw that corrections to
the kinetic term are of order $\hbar$, hence
\begin{equation}
-\nabla^2\phi+V'(\phi)=O(\hbar).
\end{equation}
These equations result in a `roll' down the potential hill of the kind
that is required by certain inflationary models
\cite{linde,hawking2,albrecht}.

The same conclusion can be reached by fixing the momentum of the fields
in the effective action \cite{evans}. This approach has some of the
advantages of the present approach, in particular it introduces
conditions in which the potential need not be convex.

The old--style effective potential has only a linear source term and
can therefore be recovered by setting $k$ to zero and solving for
$\rho$. From equations \ref{saddle} and \ref{r}, this implies that
$\rho^2+\phi^2=\mu^2/\lambda$. When $\phi^2>\mu^2/\lambda$, then there
is a single saddle point and $\rho=0$. The result is a convex potential
for $\phi$, as indeed it should be \cite{symanzik}.

\section{DOUBLE WELLS AT FINITE TEMPERATURE}

The equilibrium states of a quantum field in a canonical ensemble with
temperature $T=1/\beta$ are located at the minima of the free energy
$F=\beta\Gamma$ \cite{dolan,weinberg}, where $\Gamma$ is the finite
temperature analogue of the effective action. For a scalar field at
very high temperatures the minimum is unequivocably at the symmetric
value of the field. At low temperatures, below a critical value $T_c$,
the predominant quantum states of the scalar scalar field cause the
symmetry to be broken. Subject to a suitable solution to the convexity
problem, this should be reflected in a minimum of the effective
potential $V_\beta(\phi)$ away from the symmetric value.

Below the critical temperature it will become necessary to use the
formalism developed for the double well in the previous sections of
this paper. The change from zero temperature to the canonical ensemble
means that the spacetime should be changed to one that is periodic in
imaginary time with period $\beta$.

Homogeneous fields will be used once again, and the effective potential
defined  by
\begin{equation}
\Gamma[\phi]=\Omega V_\beta(\phi)
\end{equation}
{}From the previous result for $\Gamma$, equation \ref{main},
\begin{equation}
V_\beta=V+V^{(1)}+V^{(2)}+O(\hbar^3),
\end{equation}
where
\begin{eqnarray}
V^{(1)}=&&-\case1/2\hbar(G^{-1}-\Delta)G(x,x)
-\case1/2\hbar\Omega^{-1}\ln\det
G\\
V^{(2)}=&&\case1/86\hbar^2\lambda G(x,x)^2
\end{eqnarray}

An effective `mass' $m$ can be defined by
\begin{equation}
G^{-1}=-\nabla^2+m^2+O(\hbar)\label{pro}.
\end{equation}
(In fact $\hbar m$ has the unit of mass).
The propagator is defined by the diagrams in figure \ref{fig5}. The
ring representing $G(x,x)$ is of order $\hbar^{-2}$, but at zero
temperatures it can be absorbed into the renornalisation of the mass
and can be ignored. This is no longer the case at finite temperatures
where the rings have to be retained to keep all of the terms up to
order $\hbar$ in the propagator. Using the approximation \ref{pro}, the
effect of these rings is to shift the mass $m$,
\begin{equation}
m^2=V''(\phi)+k+3\lambda\hbar G(x,x)\label{m}.
\end{equation}

The value of $k$ is fixed by the other set of graphs \ref{fig3}, which
represent the integrals
\begin{eqnarray}
&&-\int d\mu(x')G(x,x')J_{int}(x')\nonumber\\
&&-3\lambda\hbar\int d\mu(x')G(x,x')G(x',x')\phi(x')+O(\hbar^2).
\end{eqnarray}
Multiplying by $G(x,x')^{-1}$ gives an equation for $J_{int}$,
\begin{equation}
V'(\phi)+k\phi=-3\lambda\hbar G(x,x)\phi.
\end{equation}
For the scalar field potential,
\begin{equation}
k=\mu^2-\lambda\phi^2-3\lambda\hbar G(x,x).
\end{equation}
This allows $k$ to be substituted into equation \ref{m},
\begin{equation}
m^2=2\lambda\phi^2+O(\hbar).
\end{equation}
The ring corrections have cancelled and no longer appear in the mass to
this order of approximation.

A high temperature expansion for the functional determinants and the
propagator can be used when $T>>\hbar m$ \cite{dolan,weinberg},
\begin{equation}
G(x,x)\sim\hbar^{-2}\left(\case1/{12}T^2-\case1/{6\pi}\hbar mT\right)
\end{equation}
and
\begin{equation}
\ln\det G\sim\Omega\hbar^{-4}\left(\case{\pi^2}/{45}T^4
-\case1/{12}\hbar^2m^2T^2+\case1/{6\pi}\hbar^3m^3T\right).
\end{equation}
Substituted into the effective potential, these give (dropping a
constant)
\begin{equation}
V_\beta\sim V+\case1/{12\pi}(\mu^2-\case1/4\hbar^{-1}\lambda T^2)Tm
+\case1/{16}\hbar^{-1}T^2m^2-\case1/{8\pi}Tm^3.
\end{equation}
The mass is given by $m^2=2\lambda\phi^2$. The potential is shown in
figure \ref{fig7}.

The result is different from the one we would obtain from a single
saddle point,
\begin{equation}
V_\beta\sim V+\case1/{24}\hbar^{-1}T^2m^2-\case1/{12\pi}Tm^3,
\end{equation}
with the mass given by ring corrections only,
\begin{equation}
m^2=-\mu^2+3\lambda\phi^2+\case1/{12}\hbar^{-2}T^2.
\end{equation}
However, there is agreement
on the value of the critical temperature in each case.

The new results are similar to results that can be obtained
from a `self--consistent' corrected mass \cite{linde1},
particularly with the value of $m^2=2\lambda\phi^2$. One particular
feature is the presence of a linear term. Although these are sometimes
seen in the usual approach \cite{shap}, the authors of reference
\cite{dine} have argued that they arise from a badly truncated
perturbation series. The approach presented here has been trucated
after a fixed number of loops, but it is possible to re--order the
series, if it is so desired, to truncate the series at a fixed order in
$\hbar$, $\lambda$ or $T$.

The detailed form of the potential is particularly important in the case of
a first order transition, where the field tunnells through a potential
barrier to the stable phase. The tunnelling rate is governed by a
`bounce' solution, a solution to the Riemannian (i.e. imaginary time)
equations. The bounce solution satisfies equations \ref{j} and \ref{k}
and will therefore be at a stationary point of the effective action.
The tunnelling rate depends quite sensitively on the form of the
potential.

\section{CONCLUSIONS}

The effective action is a very useful tool for the study of symmetry
breaking and phase transitions in the early universe. An attempt has
been made in this paper to construct the effective action in a
region where the potential is curving downwards as a consistent loop
expansion, both at zero or non--zero temperatures and for finite or
infinite volume systems.

We have seen that the use of quadratic sources allows a consistent
treatment of the effective action and that multiple contributions to the
path integral have to be taken into account. The corrected mass emerges
naturally rather than being introduced `by hand'. The effective action
has the form,
\begin{equation}
\Gamma[\phi]\sim I[\phi]-\case{\hbar}/2{\rm tr}(1-G\Delta)
-\case\hbar/2\ln\det G+\Gamma^{(2)},
\end{equation}
with $G$ given by the diagramatic expansion shown in figure \ref{fig5}.
The propagator $G_0$ which appears in the expansion depends upon the
number of saddle points that may contribute to the action,
\smallskip

Case I. A single saddle point, where it is possible to take the
quadratic source $K$ to be zero and $G_0$ is the inverse of the
classical fluctuation operator \ref{dint}.

Case II. Two saddle points, where $K$ is given by the vanishing of the
tadpole diagrams in figure \ref{fig4} and $G_0=(\Delta+K)^{-1}$.
\smallskip

One or the other case gives a properly defined loop expansion.
The second case applies in particular to finite
temperature systems below their critical temperature and corresponds
to what would normally be understood as the use of a `corrected mass'.

\section*{ACKNOWLEDGMENTS}

I am grateful to Janaki Balakrishnan for discussions on the subject
of this paper.

\begin{figure}
\caption{The scalar field double--well potential.\label{fig1}}
\end{figure}
\begin{figure}
\caption{The classical saddle point solution.\label{fig2}}
\end{figure}
\begin{figure}
\caption{The vertices of the shifted theory. The lines represent the
propagator $G_0$.\label{fig3}}
\end{figure}
\begin{figure}
\caption{The tadpole diagram series. Lines with circles represent the
connected propagator $G$, other circles represent connected
vertices.\label{fig4}}
\end{figure}
\begin{figure}
\caption{Iterating this series gives the diagramatic expansion of the
propagator $G$ in terms of the shifted propagator $G_0$.\label{fig5}}
\end{figure}
\begin{figure}
\caption{The effective potential $V^{(0)}(\rho,\phi)$.\label{fig6}}
\end{figure}
\begin{figure}
\caption{The finite temperature potential for a range of temperatures.
All of the temperatures are less than $T_c$.\label{fig7}}
\end{figure}


\begin{references}
\bibitem{goldstone}J. Goldstone, A. Salam and S. Weinberg, Phys. Rev.
127, 965 (1962).
\bibitem{dolan}L. Dolan and R. Jackiw, Phys. Rev. D9, 3357 (1974).
\bibitem{weinberg}S. Weinberg, Phys. Rev. Lett. 36, 294 (1976).
\bibitem{linde1}D. A. Kirzhnits and A. D. Linde, Ann. Phys. (N.Y.)
101, 195 (1976).
\bibitem{symanzik}K. Symanzik, Commun. Math. Phys. 16, 48 (1970).
\bibitem{rivers}R. Rivers, `Path integrals' (Cambridge University
Press, Cambridge 1987).
\bibitem{hawking1}S. W. Hawking and I. G. Moss, Nucl. Phys. B224, 180
(1983).
\bibitem{moss}I. G. Moss, Nucl. Phys. B238, 436 (1984).
\bibitem{cornwall}J. M. Cornwall, R. Jackiw and E. Tomboulis, Phys.
Rev. D10, 2428 (1974).
\bibitem{amelino}G. Amelino--Camelia and So--Young Pi, Phys. Rev. D47,
2356 (1993).
\bibitem{laurie}I. D. Laurie, Nucl. Phys. B301, 685 (1988).
\bibitem{cahill}K. Cahill, `A more effective potential' (University of
New Mexico preprint).
\bibitem{linde}A. D. Linde, Phys. Lett. 108B, 389 (1982).
\bibitem{hawking2}S. W. Hawking and I. G. Moss, Phys. Lett. 110B, 35
(1982).
\bibitem{albrecht}A. Albrecht and P. J. Steinhart, Phys. Rev. Lett. 48,
1220 (1982).
\bibitem{evans}M. Evans and J. G. McCarthy, Phys. Rev. D31, 1799
(1985).
\bibitem{shap} M. E. Shaposhnikov, Phys. Lett B277, 324 (1992).
\bibitem{dine}M. Dine, R. G. Leigh, P. Huet, A. D. Linde and D. Linde,
Phys. Rev. D46, 550 (1992).
\end{references}
\end{document}